\documentclass[graybox]{svmult}

\usepackage{type1cm}        
\usepackage{makeidx}         
\usepackage{graphicx}        
\usepackage{multicol}        
\usepackage[bottom]{footmisc}

\usepackage{newtxtext}       %
\usepackage{newtxmath}       

\usepackage{amssymb}
\usepackage{graphicx}
\usepackage{enumerate}
\usepackage{algorithmic}
\usepackage{algorithm}
\usepackage{url}
\usepackage{amsmath}
\usepackage{pgfplots}

\makeindex             

\begin{document}

\title*{Triclustering in Big Data Setting}
\author{Dmitry Egurnov \and Dmitry I. Ignatov \and Dmitry Tochilkin}
\authorrunning{Egurnov et al.}
\institute{Dmitry Egurnov \and Dmitry I. Ignatov \and Dmitry Tochilkin \at National Research University Higher School of Economics, Russian Federation
\email{degurnov@hse.ru}}
%
%
\maketitle

\abstract*{In this paper, we describe versions of triclustering algorithms adapted for efficient calculations in distributed environments with MapReduce model or parallelisation mechanism provided by modern programming languages. OAC-family of triclustering algorithms shows good parallelisation capabilities due to the independent processing of triples of a triadic formal context. We provide time and space complexity of the
algorithms and justify their relevance. We also compare performance gain from using a distributed system and scalability.}

\abstract{In this paper\footnote{The paper contains an extended  version of the prior work presented at the workshop on FCA in the Big Data Era held on June 25, 2019 at Frankfurt University of Applied Sciences, Frankfurt, Germany~\cite{Ignatov:2019}.}, we describe versions of triclustering algorithms adapted for efficient calculations in distributed environments with MapReduce model or parallelisation mechanism provided by modern programming languages. OAC-family of triclustering algorithms shows good parallelisation capabilities due to the independent processing of triples of a triadic formal context. We provide time and space complexity of the
algorithms and justify their relevance. We also compare performance gain from using a distributed system and scalability.}
\keywords{Formal Concept Analysis, n-ary relations, Boolean tensors, Data Mining, Big Data, MapReduce}

\section{Introduction}


Mining of multimodal patterns in $n$-ary relations or Boolean tensors is among popular topics in Data Mining and Machine Learning~\cite{Georgii:2011,Cerf:2013,Spyropoulou:2014,Ignatov:2015,Metzler:2015,Shin:2018,Henriques:2018}.  Thus, cluster analysis of multimodal data and specifically of dyadic and triadic relations is a natural extension of the idea of original clustering.
In dyadic case, biclustering methods (the term bicluster was coined in \cite{Mirkin:1996}) are used to simultaneously find subsets of objects and attributes that form homogeneous patterns of the input object-attribute data. Triclustering methods operate in triadic case, where for each object-attribute pair one assigns a set of some conditions \cite{Mirkin:2011,Ignatov:2013,Gnatyshak:2013}.
Both biclustering and triclustering algorithms are widely used in such areas as gene expression analysis \cite{Madeira:2004,Eren:2012,Zaki:2005,Li:2009,Kaytoue:2011}, recommender systems~\cite{Nanopoulos:2010,Jelassi:2013,Ignatov:2014}, social networks analysis \cite{Gnatyshak:2012}, natural language processing~\cite{Ustalov:2018},  etc. The processing of numeric multimodal data is also possible by modifications of existing approaches for mining binary relations \cite{Kaytoue:2014}. Another interesting venue closely related to attribute dependencies in object-attribute data also take place in triadic case, namely, mining of triadic association rules and implications~\cite{Ganter:2004,Missaoui:2011}.

Though there are methods that can enumerate all triclusters satisfying certain constraints \cite{Cerf:2013} (in most cases they ensure that triclusters are dense), their time complexity is rather high, as in the worst case the maximal number of triclusters is usually exponential (e.g. in case of formal triconcepts), showing that these methods are hardly scalable.
Algorithms that process big data should have at most linear time complexity (e.g., $O(|I|)$ in case of $n$-ary relation $I$) and be easily parallelisable.
In addition, especially in the case of data streams~\cite{Schweikardt18a}, the output patterns should be the results of one pass over data.

Earlier, in order to create an algorithm satisfying these requirements, we adapted a triclustering method based on prime operators (prime OAC-triclustering method) \cite{Gnatyshak:2013} and proposed its online version, which has linear time complexity; it is also one-pass and easily parallelisable \cite{Gnatyshak:2014}. However, its parallelisation is possible in different ways. For example, one can use a popular framework for commodity hardware, Map-Reduce (M/R) \cite{Rajaraman:2013}. In the past, there were several successful M/R implementations in the FCA community and other lattice-oriented domains. Thus, in \cite{Krajca:2009}, the authors adapted Close-by-One algorithm to M/R framework and showed its efficiency. In the same year, in \cite{Kudryavcev:2009}, an efficient M/R algorithm for computation of closed cube lattices was proposed. The authors of \cite{Xu:2012} demonstrated that iterative algorithms like Ganter's NextClosure can benefit from the usage of iterative M/R schemes.  

Our previous M/R implementation of the triclustering method based on prime operators was proposed in \cite{Zudin:2015} showing computational benefits on rather large datasets. M/R triclustering algorithm~\cite{Zudin:2015} is a successful distributed adaptation of the online version of prime OAC-triclustering~\cite{Gnatyshak:2014}. This method uses the MapReduce approach as means for task allocation  on computational clusters, launching the online version of prime OAC-triclustering on each reducer of the first phase.  However, due to the simplicity of this adaptation, the algorithm does not use the advantages of MapReduce to the full extent. In the first stage, all the input triples are split into the number of groups equals to the number of reducers  by means of hash-function for entities of one of the types, object, attribute, or condition, which values are used as keys. It is clear that this way of data allocation cannot guarantee uniformness in terms of group sizes. The JobTracker used in Apache Hadoop is able to evenly allocate tasks by nodes\footnote{https://wiki.apache.org/hadoop/JobTracker}. To do so, the number of tasks should be larger than the number of working nodes, which is not fulfilled in this implementation. For example, let us assume we have 10 reduce SlaveNodes; respectively, $r = 10$, and the hash function is applied to the objects (the first element in each input triple). However, due to the non-uniformity of hash-function values by modulo 10, it may happen that the set of objects will result in less than 10 different residuals during division by 10. In this case, the input triples will be distributed between parts of different sizes and processed by only a part of cluster nodes. Such cases are rather rare; it could be possible only for slicing by entities (objects, attributes, or conditions) with a small number of different elements. However, they may slow down the cluster work drastically.

The weakest link is the second stage of the algorithm. First of all, during the first stage, it finds triclusters computed for each data slice separately. Hence, they are not the final triclusters; we need to merge the obtained results.

\begin{table}[t]\label{toytri}
\caption{An example with triadic data}
\begin{center}

\begin{tabular}{ccccc}
\begin{tabular}{cccc}

& $i_1$ & $i_2$ \\
\cline{2-3}
$u_1$ &\multicolumn{1}{|c|}{$\times$} & \multicolumn{1}{|c|}{} \\
\cline{2-3}
$u_2$ &  \multicolumn{1}{|c|}{$\times$} &  \multicolumn{1}{|c|}{$\times$} \\
\cline{2-3}
$u_3$ &  \multicolumn{1}{|c|}{} &  \multicolumn{1}{|c|}{$\times$} \\
\cline{2-3}

& \multicolumn{2}{c}{$l_1$}\\

\end{tabular} & \qquad \qquad &

\begin{tabular}{cccc}

& $i_1$ & $i_2$ \\
\cline{2-3}
$u_1$ &\multicolumn{1}{|c|}{$\times$} & \multicolumn{1}{|c|}{} \\
\cline{2-3}
$u_2$ &  \multicolumn{1}{|c|}{$\times$} &  \multicolumn{1}{|c|}{$\times$} \\
\cline{2-3}
$u_3$ &  \multicolumn{1}{|c|}{$\times$} &  \multicolumn{1}{|c|}{} \\
\cline{2-3}

& \multicolumn{2}{c}{$l_2$}\\

\end{tabular}
\end{tabular}

\end{center}
\end{table}

Let us consider an example in Table~\ref{toytri} with the ternary relation on users-items-labels. Let us assume that the first mapper splits data according to their labels' component, $r = 2$, then triples containing label  $l_1$ and those related to label $l_2$ are processed on different nodes. After the first stage completion on the first node, we have tricluster $(\{u_2\}, \{i_1, i_2\}, \{l_1\})$ among the others, while the second node results in tricluster $(\{u_2\}, \{i_1, i_2\}, \{l_2\})$. It is clear that both triclusters are not complete for the whole input dataset and should be merged into $( \{u_2\}, \{i_1, i_2\}, \{l_1, l_2\})$. The second stage of the algorithm is responsible for this type of merging. However, as one can see, this merging assumes that all intermediate data should be located on the same node. In a big data setting, this allocation of all the intermediate results on a single node is a critical point for application performance.

To calculate tricluster components (or cumuli, see Section~\ref{sec:MMC}) and assemble the final triclusters from them, we need to have large data slices on the computational nodes. Moreover, during parallelisation of the algorithm, those data slices can be required simultaneously  on different nodes. Thus, to solve these problems one needs to fulfil data centralisation (all the required slices should be present at the same node simultaneously). However, it leads to the accumulation of too large parts of the data as described. Another approach is to perform data replication. In this case, the total amount of data processed on computational nodes is increased, while the data are evenly distributed in the system. The latter approach has been chosen for our updated study on multimodal clustering.

Note that experts warn the potential M/R users: ``the entire distributed-file-system milieu makes sense only when files are very large and are rarely updated in place'' \cite{Rajaraman:2013}. In this work, as in our previous study, we assume that there is a large bulk of data to process that is not coming online. As for experimental comparison, since we would like to follow an authentic M/R approach, without using the online algorithm~\cite{Gnatyshak:2014}, we implement and enhance only MapReduce schemes from the appendix in~\cite{Zudin:2015} supposed for future studies in that time. 

The rest of the paper is organised as follows: in Section~\ref{sec:OACprime}, we recall the original method and the online version of the algorithm of prime OAC-triclustering. Section~\ref{sec:extensions} is dedicated to extensions of OAC triclustering. Specifically, Subsection~\ref{sec:MMC} generalise prime OAC-triclustering for the case of multimodal data, and Subsection~\ref{sec:Multvaltric} touches the case of multi-valued context.
In Section~\ref{sec:impl} we describe implementation details. Subsection~\ref{sec:MRMC}, gives the M/R setting of the problem and the corresponding M/R version of the original algorithm with important implementation aspects. In Subsection~\ref{sec:parNOAC} we speak about parallel implementation of OAC triclustering.
Finally, in Section~\ref{sec:exp} we show the results of several experiments that demonstrate the efficiency of the M/R version of the algorithm. 

\section{Prime object-attribute-condition triclustering}\label{sec:OACprime}

Prime object-attribute-condition triclustering method (OAC-prime) based on Formal Concept Analysis \cite{Wille:1982,Ganter:1999} is an extension for the triadic case of object-attribute biclustering method \cite{Ignatov:2012a}.
Triclusters generated by this method have a similar structure as the corresponding biclusters, namely the cross-like structure of triples inside the input data cuboid (i.e. formal tricontext).

Let $\mathbb{K} = (G,M,B,I)$ be a triadic context, where $G$, $M$, $B$ are respectively the sets of objects, attributes, and conditions, and $I\subseteq G\times M\times B$ is a triadic incidence relation.
Each prime OAC-tricluster is generated by applying the following prime operators to each pair of components of some triple:
\begin{eqnarray}\label{def:primes}
    \begin{array}{c}
        (X,Y)^\prime = \{b\in B \mid (g,m,b)\in I \ {\rm for\ all}\ g\in X, m\in Y\}, \\
        (X,Z)^\prime = \{m\in M \mid (g,m,b)\in I \ {\rm for\ all}\ g\in X, b\in Z\}, \\
        (Y,Z)^\prime = \{g\in G \mid (g,m,b)\in I \ {\rm for\ all}\ m\in Y, b\in Z\},
    \end{array}
\end{eqnarray}

\noindent where $X \subseteq G$, $Y \subseteq M$, and $Z \subseteq B$.

Then the triple $T=((m,b)^\prime,(g,b)^\prime,(g,m)^\prime)$ is called \textit{prime OAC-tricluster} based on triple $(g,m,b)\in I$.
The components of tricluster are called, respectively,  \textit{tricluster extent},  \textit{tricluster intent}, and  \textit{tricluster modus}.
The triple $(g,m,b)$ is called a \textit{generating triple} of the tricluster $T$.
Density of a tricluster $T = (G_T, M_T, B_T)$ is the relation of actual number of triples in the tricluster to its size: $$\rho(T) = \frac{|G_T \times M_T \times B_T \cap I|}{|G_T||M_T||B_T|}$$ In these terms a triconcept is a tricluster with density $\rho = 1$.

\begin{figure}
\begin{center}
	\includegraphics[scale=0.7]{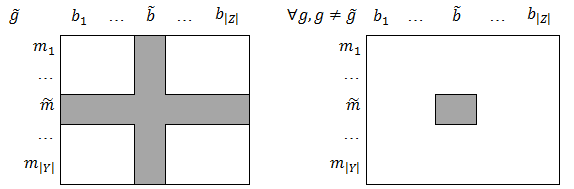}
	\caption{Structure of prime OAC-triclusters: the dense cross-like central layer containing $\tilde{g}$ (left) and the layer for an object $g$ (right) in $M \times B$ dimensions.}
	\label{prime-struct}
\end{center}
\end{figure}

Figure \ref{prime-struct} shows the structure of an OAC-tricluster $(X,Y,Z)$ based on triple $(\widetilde{g},\widetilde{m},\widetilde{b})$, triples corresponding to the gray cells are contained in the tricluster, other triples may be contained in the tricluster (cuboid) as well.

The basic algorithm for the prime OAC-triclustering method is rather straightforward (see~\cite{Gnatyshak:2013}).
First of all, for each combination of elements from each of the two sets of $\mathbb{K}$ we apply the corresponding prime operator (we call the resulting sets \noindent\emph{prime sets}).
After that, we enumerate all triples from $I$ and on each step, we must generate a tricluster based on the corresponding triple, check whether this tricluster is already contained in the tricluster set (by using hashing) and also check extra conditions.

The total time complexity of the algorithm depends on whether there is a non-zero minimal density threshold or not and on the complexity of the hashing algorithm used.
In case we use some basic hashing algorithm processing the tricluster's extent, intent and modus without a minimal density threshold, the total time complexity is $O(|G||M||B|+|I|(|G|+|M|+|B|))$ (assuming the hashing algorithm takes $O(1)$ to operate, it takes $3\times O(|G||M||B|)$ to precompute the prime sets and then $O(|G|+|M|+|B|)$ for each triple $(g,m,b) \in I$ to compute the hash); in case of a non-zero minimal density threshold, it is $O(|I||G||M||B|)$ (since computing density takes $O(|G||M||B|)$ for each tricluster generated from a triple $(g,m,b) \in I$).
The memory complexity is $O(|I|(|G|+|M|+|B|))$, as we need to keep the dictionaries with the prime sets in memory.

In online setting, for triples coming from triadic context $\mathbb{K} = (G,M,B,I)$, the user has no a priori knowledge of the elements and even cardinalities of $G$, $M$, $B$, and $I$.
At each iteration, we receive some set of triples from $I$: $J\subseteq I$.
After that, we must process $J$ and get the current version of the set of all triclusters.
It is important in this setting to consider every pair of triclusters as being different as they have different generating triples, even if their respective extents, intents, and modi are equal. Thus, any other triple can change only one of these two triclusters, making them different.

To efficiently access prime sets for their processing, the dictionaries containing the prime sets are implemented as hash-tables.

The algorithm is straightforward as well (Alg.~\ref{oac-prime-alg-online}).
It takes some set of triples ($J$), the current tricluster set ($\mathcal{T}$), and the dictionaries containing prime sets ($Primes$) as input and outputs the modified versions of the tricluster set and dictionaries.
The algorithm processes each triple $(g,m,b)$ of $J$ sequentially (line 1).
At each iteration, the algorithm modifies the corresponding prime sets (lines 2-4).

Finally, it adds a new tricluster to the tricluster set.
Note that this tricluster contains pointers to the corresponding prime sets (in the corresponding dictionaries) instead of the copies of the prime sets (line 5) which allows lowering the memory and access costs.

\begin{algorithm}
\caption{Add function for the online algorithm for prime OAC-triclustering.}
\label{oac-prime-alg-online}
    \begin{algorithmic}[1]
    	\REQUIRE $J$ is a set of triples;\\
            $\mathcal{T}=\{T=(*X,*Y,*Z)\}$ is a current set of triclusters;\\
            $PrimesOA$, $PrimesOC$, $PrimesAC$.\\
    	\ENSURE $\mathcal{T}=\{T=(*X,*Y,*Z)\}$;\\
            $PrimesOA$, $PrimesOC$, $PrimesAC$.
    	\FORALL{$(g,m,b)\in J$}
    		\STATE $PrimesOA[g,m]:=PrimesOA[g,m]\cup \{b\}$
            \STATE $PrimesOC[g,b]:=PrimesOC[g,b]\cup \{m\}$
            \STATE $PrimesAC[m,b]:=PrimesAC[m,b]\cup \{g\}$
            \STATE $\mathcal{T}:=\mathcal{T}\cup \{(\&PrimesAC[m,b],\&PrimesOC[g,b],\&PrimesOA[g,m])\}$
    	\ENDFOR
    \end{algorithmic}
\end{algorithm}

The algorithm is one-pass and its time and memory complexities are $O(|I|)$.

Duplicate elimination and selection patterns by user-specific constraints are done as post-processing to avoid patterns' loss.
%
The time complexity of the basic post-processing is $O(|I|)$  and  it does not require any additional memory. 

The algorithm can be easily parallelised by splitting the subset of triples $J$ into several subsets, processing each of them independently, and merging the resulting sets afterwards. This fact results in our previous MapReduce implementation~\cite{Zudin:2015}.

\section{Triclustering extensions}\label{sec:extensions}

\subsection{Multimodal clustering}\label{sec:MMC}

The direct extension of the prime object-attribute-condition triclustering is multimodal clustering for higher input relation arities. For the input polyadic context  $\mathbb{K}_N=(A_1,A_2,\ldots, A_N, I \subseteq A_1 \times A_2 \times \cdots \times A_N)$~\cite{Voutsadakis:02}, we introduce the notion of \emph{cumulus} for each input tuple $i=(e_1, e_2, \ldots, e_N) \in I$ and the corresponding entity $e_k$, where $k \in \{1, \ldots, N\}$ as follows:

$$cum(i,k)=\{e \mid (e_1, e_2, \ldots, e_{k-1}, e, e_{k+1}, \ldots, e_N) \in I\}.$$

The multimodal cluster generated by the tuple $i \in I$ is defined as follows:

$$\big((cum(i,1), \ldots, cum(i,N)\big) .$$

Those cumuli operators are similar to primes for pairs (or tuples) of sets Eq.~\ref{def:primes}:

$$cum(i,k)= (\{e_1\}, \{e_2\}, \ldots  , \{e_{k-1}\},\{e_{k+1}\}, \ldots, \{e_N\})^\prime.$$

However, here, they are applied to the tuples of input relation rather than to pairs (tuples) of sets.

In a certain sense, cumuli accumulate all the elements of a fixed type that are related by $I$.

As its triadic version, multimodal clustering is not greater than the number of tuples in the input relation, whereas the complete set of polyadic concepts may be exponential w.r.t. the input size~\cite{Voutsadakis:02}.

\subsection{Many-valued Triclustering}\label{sec:Multvaltric}

Another extension of prime OAC triclustering is many-valued triclustering where each triple of the incidence relation of a triadic context is associated with a certain value from an arbitrary set $W$ (by means of a valuation function $V:I\rightarrow W$). Therefore an input triadic context is changed to a many-valued triadic context $\mathbb{K_V}=(G,M,B,W,I,V)$, where for each triple $(g,m,b)$ in ternary relation $I$ between $G$, $M$, $B$, there is a unique $V(g,m,b) \in W$. 

This definition contains the explicitly given valuation function $V$ similar to the one in the definition of attribute-based information systems~\cite{Pawlak:1981}. However, we could also extend the original definition of a many-valued context in line with~\cite{Ganter:1999}. In this case, a many-valued triadic context $\mathbb{K}=(G,M,B,W,J)$ consists of sets $G$, $M$, $B$, $W$ and a quaternary relation between them $J \subseteq G \times M \times B \times W$, where the following holds:
$(g, m, b, w) \in J$ and $(g, m, b, v) \in J$ imply $w = v$. 

In what follows, we prefer the first variant since it simplifies further exposition.

For numeric values, the most common case is for $W = \mathbb{R}$. To mine triclusters in such context prime operators are modified into so-called $\delta$-operators. For a generating triple $(\tilde{g}, \tilde{m},\tilde{b})\in I$ and some parameter $\delta$:
$$(\tilde{m},\tilde{b})^\delta = \left\{g \ \mid \ (g,\tilde{m},\tilde{b})\in I \land |V(g,\tilde{m},\tilde{b}) - V(\tilde{g}, \tilde{m},\tilde{b})| \leq \delta \right\}$$
$$(\tilde{g},\tilde{b})^\delta = \left\{m \ \mid \ (\tilde{g}, m,\tilde{b})\in I \land |V(\tilde{g},m,\tilde{b}) - V(\tilde{g}, \tilde{m},\tilde{b})| \leq \delta \right\}$$
$$(\tilde{g},\tilde{m})^\delta = \left\{b \ \mid \ (\tilde{g}, \tilde{m},b)\in I \land |V(\tilde{g},\tilde{m},b) - V(\tilde{g}, \tilde{m},\tilde{b})| \leq \delta \right\}$$

This definition can still be used to mine regular triclusters, if we set $W=\left\{0,1\right\}$ and $\delta = 0$.

\section{Implementations}\label{sec:impl}

\subsection{Map-reduce-based multimodal clustering}\label{sec:MRMC}


We follow a three-stage approach here. On each stage, we sequentially run the map and reduce procedures: First map $\to$ First reduce $\to$ Second map $\to$ Second reduce $\to$ Third map $\to$ Third reduce. Each map/reduce procedure of a certain stage is executed in parallel on all the available nodes/clusters. How tasks are distributed among the nodes/clusters depends on the concrete MapReduce technology implementation (in our case, Apache Hadoop). Below, we describe the data flow between computational nodes and their processing. 

\sloppy 

1) The first map (Algorithm~\ref{map-1}) takes a set of input tuples. Each tuple $(e_1, e_2, \ldots, e_N)$ is transformed into $N$ key-value pairs: $\langle(e_2, \ldots, e_N), e_1 \rangle $,
$\langle e_1, e_3, \ldots, e_N), e_2 \rangle $, $\ldots,$ \noindent $\langle (e_1, e_2, \ldots, e_{N-1}), e_N \rangle $. The resulting pairs are passed to the further step.

\begin{algorithm}
\caption{Distributed Multimodal clustering: First Map}
\label{map-1}
    \begin{algorithmic}[1]
    	\REQUIRE $I$ is a set of tuples of length $N$ each\\
	   	\ENSURE  $\langle subrelation, entity \rangle$ pairs.\\
    	\FORALL{$(e_1, e_2, \ldots, e_N) \in I$}
    	    \FORALL{$k \in \{1,\ldots,N\}$}
                \STATE $subrelation := (e_1, \ldots, e_{k-1}, e_{k+1} ,\ldots,\ e_N)$
                \STATE \textbf{emit} $\langle subrelation, e_k \rangle$
    	    \ENDFOR
    	\ENDFOR
    	
    \end{algorithmic}
\end{algorithm}

2) The first reduce (Algorithm~\ref{reduce-1}) receives all the accumulated values of each key. Thus, for each $(e_1, \ldots,\ e_N) \in I$ and the context entity type $k \in \{1, 2, \ldots, N\}$, we compute the cumulus  $(e_1, \ldots, e_{k-1}, e_{k+1} ,\ldots,\ e_N)^\prime$. The values are passed to the next MapReduce stage with the key $(e_1, \ldots, e_{k-1}, e_{k+1} ,\ldots,\ e_N)$.

\begin{algorithm}
\caption{Distributed Multimodal clustering: First Reduce}
\label{reduce-1}
    \begin{algorithmic}[1]
    	\REQUIRE key-value pairs $\langle subrelation, entities \ \{e_k^1, \ldots, e_k^L \}\rangle$ \\
	   	\ENSURE  $\langle subrelation, cumulus \rangle$\\
	    \STATE 	cumulus:=\{\}
	   	\FORALL{$e_k \in \{e_k^1, \ldots, e_k^L\}$}
                \STATE $cumulus := cumulus \cup \{e_k\}$
                \STATE \textbf{emit} $\langle subrelation, cumulus \rangle$
    	    \ENDFOR
	   	
    \end{algorithmic}
\end{algorithm}

3) Second map (Algorithm~\ref{map-2}). All the received keys are transformed into the original relations and passed to the second reduce procedure with unchanged values.

\begin{algorithm}
\caption{Distributed Multimodal clustering: Second Map}
\label{map-2}
    \begin{algorithmic}[1]
    	\REQUIRE $\langle subrelation, cumulus \rangle$, where $subrelation=(e_1, \ldots, e_{k-1}, e_{k+1} ,\ldots,\ e_N)$\\
	   	\ENSURE  $\langle generating\_relation, cumulus \rangle$ pairs.\\
    	\FORALL{$e_k \in cumulus$}
            \STATE $generating\_relation := (e_1, \ldots, e_{k-1}, e_{k}, e_{k+1} ,\ldots,\ e_N)$
            \STATE \textbf{emit} $\langle generating\_relation, cumulus \rangle$
    	\ENDFOR
    \end{algorithmic}
\end{algorithm}

4) Second reduce (Algorithm~\ref{reduce-2}). All the cumuli obtained for each input tuple of the original relation $I$ are reduced to a single set. At this stage, we obtain all the original tuples and generated multimodal clusters. These clusters are presented as tuples of cumuli for respective entity types. All the obtained pairs  $\langle generating\_relation, multimodal\_cluster \rangle$ are passed to the next stage.

\begin{algorithm}
\caption{Distributed Multimodal clustering: Second Reduce}
\label{reduce-2}
    \begin{algorithmic}[1]
    	\REQUIRE $\langle generating\_relation, cumuli \ \{A_1, A_2, \cdots, A_N\} \rangle$\\
	   	\ENSURE  $\langle  generating\_relation, multimodal\_cluster \rangle$ pairs\\
	   	\STATE $multimodal\_cluster: = (A_1, A_2, \cdots, A_N)$
	   	\STATE \textbf{emit} $\langle generating\_relation, multimodal\_cluster \rangle$
    \end{algorithmic}
\end{algorithm}

5) Third map (Algorithm~\ref{map-3}). The task of the third MapReduce stage is duplicate elimination and filtration by density threshold. It is beneficial to implement within the reduce step, but to do so each obtained key-value pair $\langle generating\_relation,$ $ multimodal\_cluster \rangle$ should be passed further as follows  $\langle  multimodal\_cluster,$ $generating\_relation \rangle$.

\begin{algorithm}
\caption{Distributed Multimodal clustering: Third Map}
\label{map-3}
    \begin{algorithmic}[1]
    	\REQUIRE $\langle generating\_relation, multimodal\_cluster \rangle$
        \STATE \textbf{emit} $\langle multimodal\_cluster, generating\_relation \rangle$
    \end{algorithmic}
\end{algorithm}

6) Third reduce (Algorithm~\ref{reduce-3}). For each input multimodal cluster and its generating tuples, it is possible to directly compute density. All the unique clusters will be stored.

\begin{algorithm}
\caption{Distributed Multimodal clustering: Third Reduce}
\label{reduce-3}
    \begin{algorithmic}[1]
    	\REQUIRE  $\langle  multimodal\_cluster, generating\_relations \ \{r_1, r_2 \ldots, r_M\} \rangle$
    		   	\IF{$\dfrac{| \{r_1, r_2 \ldots, r_M\}|}{vol(multimodal\_cluster)}  \geq \theta$}
        \STATE \textbf{store} $\langle multimodal\_cluster \rangle$
          \ENDIF
    \end{algorithmic}
\end{algorithm}

The time and memory complexities are provided assuming that the worst-case scenario corresponds to the absolutely dense cuboid, i.e. polyadic context $\mathbb{K}_N=(A_1,A_2,\ldots, A_N, A_1 \times A_2 \times \cdots \times A_N)$. Thus, after careful analysis, the worst-case time complexity of the proposed three-stage algorithm is $O(|I|\sum_{j=1}^{N}|A_j|)$. Not surprisingly it has the same worst-case memory complexity since the stored and passed the maximal number of multimodal clusters is $|I|$, and the size of each of them is not greater $\sum_{j=1}^{N}|A_j|$. However, from an implementation point of view, since HDFS has default replication factor 3, those data elements are copied thrice to fulfil fault-tolerance.

\subsection{Implementation aspects and used technologies}

The application\footnote{\url{https://github.com/kostarion/multimodal-clustering-hadoop}} has been implemented in Java and as distributed computation framework we use Apache Hadoop\footnote{ \url{https://hadoop.apache.org/}}.

We have used many other technologies: Apache Maven (framework for automatic project assembling), Apache Commons (for work with extended Java collections), Jackson JSON (open-source library for the transformation of object-oriented representation of an object like tricluster to string), TypeTools (for real-time type resolution of inbound and outbound key-value pairs),  etc. 

To provide the reader with basic information on the most important classes for  M/R implementation, let us shortly describe them below.

\noindent\emph{Entity.} This is a basic abstraction for an element of a certain type. Each entity is defined by its type index from 0 to $n$-1, where $n$ is the arity of the input formal context. An entity value is a string that needs to be kept during the program execution. This class inherits Writable interface for storing its objects in temporary and permanent Hadoop files. This is a mandatory requirement  for all classes that pass or take their objects as keys and values of the map and reduce methods.

\noindent\emph{Tuple.} This class implements a representation of relation. Each object of the class Tuple contains the list of objects of Entity class and the arity of its data given by its numeric value. This class implements interface WritableComparable$<$Tuple$>$ to make it possible to use an object  class Tuple as a key. The interface is similar to Writable, however, one needs to define comparison function to use in the key sorting phase.

\noindent\emph{Cumulus.} This is an abstraction of cumulus, introduced earlier. It contains the list of string values and the index of a respective entity. It also implements the following interfaces: WritableComparable$<$Cumulus$>$ for using cumulus as a key and  Iterable$<$String$>$ for iteration by its values.

\noindent\emph{FormalConcept.} This is an abstraction of both formal concepts and multimodal clusters, it contains the list of cumuli and implements interface Writable.

The process-like M/R classes are summarised below.

\noindent\emph{FirstMapper, SecondMapper, ThirdMapper.} These are the classes that extend class Mapper$<>$ of the Hadoop MapReduce library by respective mapping function from Subsection~\ref{sec:MRMC}.

\noindent\emph{FirstReducer, SecondReducer, ThirdReducer.} These classes extend class Reducer$<>$ of the Hadoop MapReduce library for fulfilling Algorithms~\ref{map-1},\ref{map-2},\ref{map-3}.

\noindent\emph{TextCumulusInputFormat, TextCumulusOutputFormat.}
These classes implement reading and writing for objects of Cumulus class; they also inherit RecordReader and RecordWriter interfaces, respectively. They are required to exchange results between different MapReduce phases within one MapReduce program.

\noindent\emph{JobConfigurator.} This is the class for setting configuration of a single MapReduce stage. It defines the classes of input/output/intermediate keys and values of the mapper and reducer as well as formats of input and output data. 

\noindent\emph{App.} This class is responsible for launching the application and chaining of M/R stages. 

\subsection{Parallel many-valued triclustering}\label{sec:parNOAC}

A generic algorithm for OAC triclustering is described below

\begin{algorithm}
\caption{General algorithm for OAC triclustering}
\label{generalOAC}
    \begin{algorithmic}[1]
    	\REQUIRE Context $\mathbb{K}=(G,M,B,I)$
	   	\ENSURE  Set of triclusters $\mathcal{T}$
	   	\STATE $\mathcal{T} := \emptyset$
    	\FORALL{$(g,m,b) \in I$}
            \STATE $oSet := applyPrimeOperator(m,b)$
            \STATE $aSet := applyPrimeOperator(g,b)$
            \STATE $cSet := applyPrimeOperator(g,m)$
            \STATE $tricluster := (oSet, aSet, cSet)$
            \IF{$tricluster$ is valid}
                \STATE $Add(\mathcal{T}, tricluster)$
            \ENDIF
    	\ENDFOR
        \RETURN $\mathcal{T}$
    \end{algorithmic}
\end{algorithm}

To get a specific version of the algorithm one only needs to add an appropriate implementation of the prime operator and optional validity check. A tricluster mined from one triple does not depend on triclusters mined from other triples, so, in case of parallel implementation, each triple is processed in an individual thread.

We used $\delta$-operators defined in \ref{sec:Multvaltric}, minimal density, and minimal cardinality (w.r.t. to every dimension) constraints~\cite{Egurnov2017}.

\section{Experiments}\label{sec:exp}

Two series of experiments have been conducted in order to test the application on the synthetic contexts and real-world datasets with a moderate and large number of triples in each.
In each experiment, both versions of the OAC-triclustering algorithm have been used to extract triclusters from a given context.
Only online and M/R versions of OAC-triclustering algorithm have managed to result in patterns for large contexts since the computation time of the compared algorithms was too high ($>$3000 s).
To evaluate the runtime more carefully, for each context the average result of 5 runs of the algorithms has been recorded.

\subsection{Datasets}

\noindent\emph{Synthetic datasets.} The following synthetic datasets were generated.

\noindent The dense context $\mathbb{K}_1=(G,M,B,I)$, where $G=M=B=\{1,\ldots, 60\}$ and $I=G \times M \times B \setminus \{(g,m,b) \in I \mid g=m=b \}$. In total,  $60^3$ - 60 =215,940 triples.

\noindent The context of three non-overlapped cuboids $\mathbb{K}_2=(G_1 \sqcup G_2 \sqcup G_3,M_1 \sqcup M_2 \sqcup M_3,B_1 \sqcup B_2 \sqcup B_3, I)$, where $I = (G_1 \times M_1 \times B_1) \cup (G_2 \times M_2 \times B_2) \cup (G_3 \times M_3 \times B_3$). In total, $3 \cdot 50^3 = 375,000$ triples.

\noindent The context $\mathbb{K}_3=(A_1,A_2,A_3,A_4, A_1 \times A_2 \times A_3 \times A_4 )$ is a dense fourth dimensional cuboid with $|A_1|=|A_2|=|A_3|=|A_4|=30$ containing $30^4 = 810,000$ triples.

These tests have sense since in M/R setting due to the tuples can be (partially) repeated, e.g., because of M/R task failures on some nodes (i.e. restarting processing of some key-value pairs). Even though the third dataset does not result in $3^{min(|A_1|,|A_2|,|A_3|,|A_4|)}$ formal triconcepts, the worst-case for formal triconcepts generation in terms of the number of patterns, this is an example of the worst-case scenario for the reducers since the input has its maximal size w.r.t. to the size of $A_i$-s and the number of duplicates. In fact, our algorithm correctly assembles the only one tricluster $(A_1,A_2,A_3,A_4)$. 

\noindent\emph{IMDB.} This dataset consists of the 250 best movies from the Internet Movie Database based on user reviews.

The following triadic context is composed: the set of objects consists of movie names, the set of attributes (tags), the set of conditions (genres), and each triple of the ternary relation means that the given movie has the given genre and is assigned the given tag. In total, there are 3,818 triples. 

\noindent\emph{Movielens.} The dataset contains 1,000,000 tuples that relate 6,040 users, 3,952 movies, ratings, and timestamps, where ratings are made on a 5-star scale~\cite{Harper:2016}.

\noindent\emph{Bibsonomy.} Finally, a sample of the data of bibsonomy.org from ECML PKDD discovery challenge 2008 has been used.

This website allows users to share bookmarks and lists of literature and tag them.
For the tests the following triadic context has been prepared: the set of objects consists of users, the set of attributes (tags), the set of conditions (bookmarks), and a triple of the ternary relation means that the given user has assigned the given tag to the given bookmark.

Table~\ref{tbl:main-cont} contains the summary of IMDB and Bibsonomy contexts.

\begin{table}[!htb]
\caption{Tricontexts based of real data systems for the experiments}
\label{tbl:main-cont}
	\begin{center}
		\begin{tabular}{|c|r|r|r|r|l|}
			\hline
			Context & $|G|$ & $|M|$ & $|B|$ & \# triples & Density \\
			\hline
			IMDB & 250 & 795 & 22 & 3,818 & 0.00087 \\
			BibSonomy & 2,337 & 67,464 & 28,920 & 816,197 & $1.8 \cdot 10^{-7}$ \\
			\hline
		\end{tabular}
	\end{center}
\end{table}

Due to a number of requests, we would like to show small excerpts of the input data we used (in this section) and a few examples of the outputted patterns (in the next section).

\begin{svgraybox}
\textbf{Input data example.} The input file for the Top-250 IMDB dataset comprises triples in lines with tab characters as separators.

\

One Flew Over the Cuckoo's Nest (1975) \quad Nurse  \quad Drama

One Flew Over the Cuckoo's Nest (1975) \quad Patient \quad Drama

One Flew Over the Cuckoo's Nest (1975) \quad Asylum \quad Drama

One Flew Over the Cuckoo's Nest (1975) \quad Rebel \quad Drama

One Flew Over the Cuckoo's Nest (1975) \quad Basketball \quad Drama

Star Wars V:  The Empire Strikes Back (1980) \quad Princess \quad Action

Star Wars V: The Empire Strikes Back (1980) \quad Princess \quad Adventure

Star Wars V: The Empire Strikes Back (1980) \quad  Princess \quad Sci-Fi

\ldots

\end{svgraybox}

\subsection{Results}

The experiments have been conducted on the computer Intel\textregistered Core(TM) i5-2450M CPU @ 2.50GHz, 4Gb RAM (typical commodity hardware) in the emulation mode, when Hadoop cluster contains only one node and operates locally and sequentially. By time execution results one can estimate the performance in a real distributed environment assuming that each node workload is (roughly) the same.

\begin{table}[!htb]
\caption{Three-stage MapReduce multimodal clustering time, ms}\label{tbl:res1}
\small
\begin{center}
\begin{tabular}{|l|c|c|c|c|c|c|}
\hline
Method & IMDB & MovieLens100k & $\mathbb{K}_1$ & $\mathbb{K}_2$ & $\mathbb{K}_3$ \\
\hline 
Online OAC prime clustering & 368 & 16,298 & 96,990 & 185,072 & 643,978 \\
MapReduce multimodal clustering & 7,124 & 14,582 & 37,572 & 61,367 & 102,699 \\
\hline
\end{tabular}
\end{center}
\end{table}

\begin{table}[!htb]
\caption{Three-stage MapReduce multimodal clustering time, ms}\label{tbl:res2}
\small
\begin{center}
\begin{tabular}{|l|c| c|c|c|c|c|}
\hline
Dataset & Online  	&	M/R total &	 \multicolumn{3}{c|}{MapReduce stages} & \# clusters\\
 \cline{4-6} & OAC Prime &	 &	1st  &	2nd   &	3rd  & \\
\hline
MovieLens100k & 89,931 &	16,348 &	8,724 &	5,292 &	2,332 &	89,932	\\
MovieLens250k &	225,242 & 42,708 &	10,075 &	20,338 &	12,295 &	225,251	\\
MovieLens500k & 461,198	& 94,701 &	15,016 &	46,300 &	33,384 &	461,238	\\
MovieLens1M &  958,345 &	217,694 & 28,027 & 114,221 & 74,446 & 942,757\\
\hline
Bibsonomy	& $>$ 6 hours & 3,651,072	& 19,117 &	1,972,135	& 1,659,820	& 486,221\\
$(\approx$800k triples) & & $(\approx$1 hour) & & & & \\
\hline
\end{tabular}
\end{center}
\end{table}

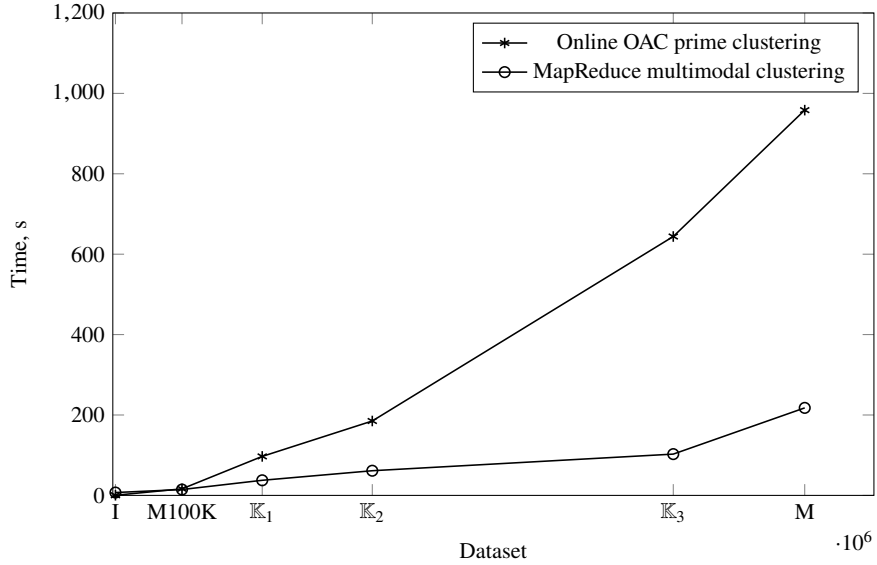
\begin{figure}[t!]
\begin{tikzpicture}
    \begin{axis}[width=\linewidth, height=8cm, xmin=0, xmax=1100000, xlabel={\footnotesize Dataset}, ylabel={\footnotesize Time, s},
          xticklabels=  {I, M100K , $\mathbb{K}_1$ , $\mathbb{K}_2$ , $\mathbb{K}_3$, M},
          xtick=        {3818,100000,215940,375000,810000,1000000}, ymin=0, ymax=1200,
          x tick label style={font=\footnotesize}, y tick label style={font=\footnotesize}
          ]
      \addplot[line width=.6pt,black,solid,mark=asterisk,mark repeat=1] table {
      3818 .368
      100000 16.298
      215940 96.990
      375000 185.072
      810000 643.978
      1000000 958.345
      };
      \addplot[line width=.6pt,black,solid,mark=o,mark repeat=1] table {
      3818 7.124 
      100000 14.582
      215940 37.572
      375000 61.367
      810000 102.699
      1000000 217.694
      };
      \addlegendentry{Online OAC prime clustering}
      \addlegendentry{MapReduce multimodal clustering}
    \end{axis}
\end{tikzpicture}
\caption{Performance curves for six datasets: I stands for IMDB dataset with 3,818 triples, M100K -- MovieLens dataset with 100K tuples, M -- MovieLens dataset with 1M tuples }\label{fig:perf_curves}
\end{figure}

In Tables~\ref{tbl:res1} and \ref{tbl:res2}  we summarise the results of performed tests. It is clear that on average our application has a smaller execution time than its competitor, the online version of OAC-triclustering. If we compare the implemented program with its original online version, the results are worse for the not that big and sparse dataset as IMDB. It is the consequence of the fact that the application architecture aimed at processing large amounts of data; in particular, it is implemented in three stages with time-consuming communication. Launching and stopping Apache Hadoop, data writing, and passing between Map and Reduce steps in both stages requires substantial time, that is why for not that big datasets when execution time is comparable with time for infrastructure management, time performance is not perfect. However, with data size increase the relative performance is growing up to five-six times (see Fig.~\ref{fig:perf_curves}).
Thus, the last test for BibSonomy data has been successfully passed, but the competitor  was not able to finish it within one hour. As for the M/R stages, the most time-consuming phases are the 2nd and 3rd stages.

\begin{svgraybox}
\textbf{Output triclusters example.} The output triclusters are stored in a format, similar to JSON: the sets of entities (modalities) are given in curly brackets separated with commas. Both sets of entities and triclusters start with a new line.

Thus, for the modality of movies, in case of IMDB data, we can see two remaining modalities as keywords and genres.

\begin{verbatim}
{
{Apocalypse Now (1979), Forrest Gump (1994),
Full Metal Jacket (1987), Platoon (1986)}
{Vietnam}
{Drama, Action}
}
{
{Toy Story (1995), Toy Story 2 (1999)}
{Toy, Friend}
{Animation, Adventure, Comedy, Family, Fantasy}
}
{
{Star Wars: Episode V – The Empire Strikes Back (1980),
WALL-E (2008), Toy Story 2 (1999)}
{Rescue}
{Animation, Adventure}
}
{
{Into the Wild (2007), The Gold Rush (1925)}
{Love, Alaska}
{Adventure}
}
...
\end{verbatim}

\end{svgraybox}

\section{Experiments with parallelisation}

In the additional set of experiments, we investigated parallelisation as another way of improving the performance of our triclustering algorithms. Modern computers support multi-threading and have several processing cores. However one needs special instructions and thread-safe data structures to produce an efficient parallel algorithm.

We used a dataset on semantic tri-frames featured in~\cite{Ustalov:2018}. These triples represent semantic frames extracted from FrameNet1.7~\cite{Baker:1998} and each triple is accompanied by a frequency from DepCC dataset~\cite{PANCHENKO18}. The total count of triples reached 100 thousand. 
The data was processed with the NOAC algorithm~\cite{Egurnov2017} implemented in C\# .NetFramework 4.5~\cite{Egurnov2019} and  modified for parallel computation with Parallel library, namely each triple from the context is processed in a separate thread. 
We ran two series of experiments with different algorithm parameters and measured execution time against the number of processed triples and also provided the number of extracted triclusters. In Table~\ref{tbl:res3}, for example, the record NOAC(100, 0.8, 2) 1k means that $\delta=100$, $\rho_{min}=0.8$, $minsup=2$ (for each dimension) and 1,000 triples are being processed.
All additional experiments were conducted on an Intel\textregistered Core(TM) i7-8750H CPU @ 2.20GHz, 16Gb RAM.

\begin{table}[!htb]
\caption{NOAC. Regular ans parallel version}\label{tbl:res3}
\small
\begin{center}
\begin{tabular}{|l|c|c|c|}
\hline
Experiment & Time, ms (regular) & Time, ms (parallel) & \# Triclusters \\
\hline
NOAC(100, 0.8, 2) 1k    & 109       & 117       & 0 \\
NOAC(100, 0.8, 2) 10k   & 5,025     & 3,642     & 3 \\
NOAC(100, 0.8, 2) 20k   & 16,825    & 11,759    & 20 \\
NOAC(100, 0.8, 2) 30k   & 33,067    & 22,519    & 52 \\
NOAC(100, 0.8, 2) 40k   & 56,878    & 36,994    & 92 \\
NOAC(100, 0.8, 2) 50k   & 80,095    & 52,322    & 116 \\
NOAC(100, 0.8, 2) 60k   & 102,092   & 67,748    & 145 \\
NOAC(100, 0.8, 2) 70k   & 133,974   & 87,956    & 160 \\
NOAC(100, 0.8, 2) 80k   & 175,597   & 110,044   & 201 \\
NOAC(100, 0.8, 2) 90k   & 223,932   & 135,268   & 223 \\
NOAC(100, 0.8, 2) 100k  & 268,021   & 157,073   & 254 \\
\hline
NOAC(100, 0.5, 0) 1k    & 110       & 169       & 803 \\
NOAC(100, 0.5, 0) 10k   & 5,121     & 3,681     & 4942 \\
NOAC(100, 0.5, 0) 50k   & 82,130    & 52,558    & 14214 \\
NOAC(100, 0.5, 0) 100k  & 268,128   & 159,333   & 23134 \\
\hline
\end{tabular}
\end{center}
\end{table}

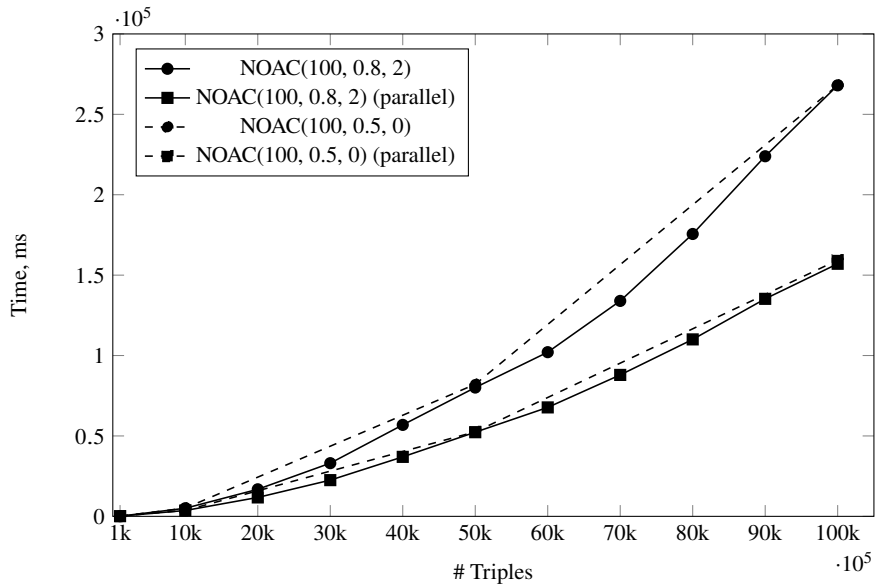
\begin{figure}[t!]
\begin{tikzpicture}
    \begin{axis}[
            width=\linewidth, 
            height=8cm, 
            xmin=0, 
            xmax=105000,
            xlabel={\footnotesize \# Triples}, 
            ylabel={\footnotesize Time, ms},
            xticklabels={1k, 10k, 20k, 30k, 40k, 50k, 60k, 70k, 80k, 90k, 100k},
            xtick={1000,10000,20000,30000,40000,50000,60000,70000,80000,90000,100000}, 
            ymin=0,
            ymax=300000,
            x tick label style={font=\footnotesize}, 
            y tick label style={font=\footnotesize},
            legend pos=north west
        ]
        \addlegendentry{NOAC(100, 0.8, 2)}
        \addplot[line width=.6pt,black,solid,mark=*,mark repeat=1] table {
            1000     109  
            10000    5025  
            20000    16825 
            30000    33067 
            40000    56878 
            50000    80095 
            60000    102092
            70000    133974
            80000    175597
            90000    223932
            100000   268021
        };
        \addlegendentry{NOAC(100, 0.8, 2) (parallel)}
        \addplot[line width=.6pt,black,solid,mark=square*,mark repeat=1] table {
            1000   117    
            10000  3642  
            20000  11759 
            30000  22519 
            40000  36994 
            50000  52322 
            60000  67748 
            70000  87956 
            80000  110044
            90000  135268
            100000 157073
        };
        \addlegendentry{NOAC(100, 0.5, 0)}
        \addplot[line width=.6pt,black,dashed,mark=*,mark repeat=1] table {
            1000     110   
            10000    5121  
            50000    82130 
            100000   268128
        };
        \addlegendentry{NOAC(100, 0.5, 0) (parallel)}
        \addplot[line width=.6pt,black,dashed,mark=square*,mark repeat=1] table {
            1000     169  
            10000   3681  
            50000   52558 
            100000  159333
        };
    \end{axis}
\end{tikzpicture}
\caption{Performance curves for parallelisation experiments} \label{fig:perf_curves_parallel}
\end{figure}

These experiments show that performance noticeably benefits from parallelisation. The execution time of the parallel algorithm is on average 35\% lower. Another interesting outcome is that execution time does not depend on the algorithm parameters, which only change the number of extracted triclusters.

\section{Conclusion}

In this paper, we have presented a map-reduce version of the multimodal clustering algorithm, which extends triclustering approaches and copes with bottlenecks of the earlier M/R triclustering version~\cite{Zudin:2015}. We have shown that the proposed algorithm is efficient from both theoretical and practical points of view. This is a variant of map-reduce based algorithm where the reducer exploits composite keys directly (see also  Appendix section~\cite{Zudin:2015}). However, despite the step towards Big Data technologies, a proper comparison of the proposed multimodal clustering and noise-tolerant patterns in $n$-ary relations by DataPeeler and its descendants~\cite{Cerf:2013} is not yet conducted (including MapReduce setting). 

Two of the most challenging problems in OAC-triclustering are approximate triclustering density estimation (e.g., employing the Monte Carlo approach) and duplicate elimination (the same triclusters can be generated by different generating triples). In MapReduce setting, both procedures are implemented as separate MapReduce stages, while for online triclustering generation it is better to avoid duplicate generation, and the tricluster density should be computed both approximately to maintain appropriate time-complexity and iteratively with minimal updates for each incoming triple. 

Further development of the proposed triclustering methods for large datasets is possible with Apache Spark\footnote{https://spark.apache.org/}~\cite{Zaharia:2016}. 

We also examined parallelisation possibilities (namely, multi-thread and milti-core advantages in modern C\#) of the numeric  triclustering algorithm NOAC and showed that it may significantly improve performance.

\subsubsection*{Acknowledgements.} The study was implemented in the framework of the Basic Research Program at the National Research University Higher School of Economics (Sections 2 and 5), and funded by the Russian Academic Excellence Project '5-100'. The second author was also supported by the Russian Science Foundation (Section 1, 3, and 4) under grant 17-11-01294. The authors would like to thank Dmitry Gnatyshak and Sergey Zudin for their earlier work on incremental and MapReduce-based triclustering implementations, respectively, anonymous reviewers as well as Yuri Kudriavtsev from PM-Square and Dominik Slezak from Infobright and Warsaw University for their encouragement given to our studies of M/R technologies.

\bibliographystyle{plain}
\bibliography{ref}

\begin{thebibliography}{10}

\bibitem{Baker:1998}
Collin~F. Baker, Charles~J. Fillmore, and John~B. Lowe.
\newblock The berkeley framenet project.
\newblock In {\em Proceedings of the 36th Annual Meeting of the Association for
  Computational Linguistics and 17th International Conference on Computational
  Linguistics - Volume 1}, ACL '98/COLING '98, pages 86--90, Stroudsburg, PA,
  USA, 1998. Association for Computational Linguistics.

\bibitem{Cerf:2013}
Lo\"{\i}c Cerf, J{\'e}r{\'e}my Besson, Kim-Ngan Nguyen, and Jean-Fran\c{c}ois
  Boulicaut.
\newblock Closed and noise-tolerant patterns in n-ary relations.
\newblock {\em Data Min. Knowl. Discov.}, 26(3):574--619, 2013.

\bibitem{Egurnov2017}
Dmitrii Egurnov, Dmitry Ignatov, and Engelbert~Mephu Nguifo.
\newblock Mining triclusters of similar values in triadic real-valued contexts.
\newblock In {\em 14th International Conference on Formal Concept
  Analysis-Supplementary Proceedings}, pages 31--47, 2017.

\bibitem{Egurnov2019}
Dmitrii Egurnov and Dmitry~I. Ignatov.
\newblock Triclustring toolbox.
\newblock In {\em Supplementary Proceedings of {ICFCA} 2019 Conference and
  Workshops, Frankfurt, Germany, June 25-28, 2019}, pages 65--69, 2019.

\bibitem{Eren:2012}
Kemal Eren, Mehmet Deveci, Onur Kucuktunc, and {Catalyurek, Umit V.}
\newblock A comparative analysis of biclustering algorithms for gene expression
  data.
\newblock {\em Briefings in Bioinform.}, 2012.

\bibitem{Ganter:2004}
Bernhard Ganter, Peter~A. Grigoriev, Sergei~O. Kuznetsov, and Mikhail~V.
  Samokhin.
\newblock Concept-based data mining with scaled labeled graphs.
\newblock In {\em ICCS}, pages 94--108, 2004.

\bibitem{Ganter:1999}
Bernhard Ganter and Rudolf Wille.
\newblock {\em {Formal Concept Analysis: Mathematical Foundations}}.
\newblock Springer-Verlag New York, Inc., Secaucus, NJ, USA, 1st edition, 1999.

\bibitem{Georgii:2011}
Elisabeth Georgii, Koji Tsuda, and Bernhard Sch\"{o}lkopf.
\newblock Multi-way set enumeration in weight tensors.
\newblock {\em Machine Learning}, 82(2):123--155, 2011.

\bibitem{Gnatyshak:2013}
Dmitry~V. Gnatyshak, Dmitry~I. Ignatov, and Sergei~O. Kuznetsov.
\newblock From triadic {FCA} to triclustering: Experimental comparison of some
  triclustering algorithms.
\newblock In {\em CLA}, pages 249--260, 2013.

\bibitem{Gnatyshak:2014}
Dmitry~V. Gnatyshak, Dmitry~I. Ignatov, Sergei~O. Kuznetsov, and Lhouari
  Nourine.
\newblock A one-pass triclustering approach: Is there any room for big data?
\newblock In {\em CLA 2014}, 2014.

\bibitem{Gnatyshak:2012}
Dmitry~V. Gnatyshak, Dmitry~I. Ignatov, Alexander~V. Semenov, and Jonas
  Poelmans.
\newblock Gaining insight in social networks with biclustering and
  triclustering.
\newblock In {\em BIR}, volume 128 of {\em Lecture Notes in Business
  Information Processing}, pages 162--171. Springer, 2012.

\bibitem{Harper:2016}
F.~Maxwell Harper and Joseph~A. Konstan.
\newblock The movielens datasets: History and context.
\newblock {\em TiiS}, 5(4):19:1--19:19, 2016.

\bibitem{Henriques:2018}
Rui Henriques and Sara~C. Madeira.
\newblock Triclustering algorithms for three-dimensional data analysis: A
  comprehensive survey.
\newblock {\em ACM Comput. Surv.}, 51(5):95:1--95:43, September 2018.

\bibitem{Ignatov:2015}
Dmitry~I. Ignatov, Dmitry~V. Gnatyshak, Sergei~O. Kuznetsov, and Boris Mirkin.
\newblock Triadic formal concept analysis and triclustering: searching for
  optimal patterns.
\newblock {\em Machine Learning}, pages 1--32, 2015.

\bibitem{Ignatov:2012a}
Dmitry~I. Ignatov, Sergei~O. Kuznetsov, and Jonas Poelmans.
\newblock Concept-based biclustering for internet advertisement.
\newblock In {\em ICDM Workshops}, pages 123--130. IEEE Computer Society, 2012.

\bibitem{Ignatov:2013}
Dmitry~I. Ignatov, Sergei~O. Kuznetsov, Jonas Poelmans, and Leonid~E. Zhukov.
\newblock Can triconcepts become triclusters?
\newblock {\em International Journal of General Systems}, 42(6):572--593, 2013.

\bibitem{Ignatov:2014}
Dmitry~I. Ignatov, Elena Nenova, Natalia Konstantinova, and Andrey~V.
  Konstantinov.
\newblock {Boolean Matrix Factorisation for Collaborative Filtering: An
  FCA-Based Approach}.
\newblock In {\em AIMSA 2014, Varna, Bulgaria, Proceedings}, volume LNCS 8722,
  pages 47--58, 2014.

\bibitem{Ignatov:2019}
Dmitry~I. Ignatov, Dmitry Tochilkin, and Dmitry Egurnov.
\newblock Multimodal clustering of boolean tensors on mapreduce: Experiments
  revisited.
\newblock In {\em Supplementary Proceedings of {ICFCA} 2019 Conference and
  Workshops, Frankfurt, Germany, June 25-28, 2019}, pages 137--151, 2019.

\bibitem{Jelassi:2013}
Mohamed~Nader Jelassi, Sadok~Ben Yahia, and Engelbert~Mephu Nguifo.
\newblock A personalized recommender system based on users' information in
  folksonomies.
\newblock In Leslie Carr and et~al., editors, {\em WWW (Companion Volume)},
  pages 1215--1224. ACM, 2013.

\bibitem{Kaytoue:2014}
Mehdi Kaytoue, Sergei~O. Kuznetsov, Juraj Macko, and Amedeo Napoli.
\newblock Biclustering meets triadic concept analysis.
\newblock {\em Ann. Math. Artif. Intell.}, 70(1-2):55--79, 2014.

\bibitem{Kaytoue:2011}
Mehdi Kaytoue, Sergei~O. Kuznetsov, Amedeo Napoli, and S{\'{e}}bastien
  Duplessis.
\newblock Mining gene expression data with pattern structures in formal concept
  analysis.
\newblock {\em Inf. Sci.}, 181(10):1989--2001, 2011.

\bibitem{Krajca:2009}
Petr Krajca and Vilem Vychodil.
\newblock Distributed algorithm for computing formal concepts using map-reduce
  framework.
\newblock In {\em N. Adams et al. (Eds.): IDA 2009}, volume LNCS 5772, pages
  333--344, 2009.

\bibitem{Kudryavcev:2009}
Sergey Kuznecov and Yury Kudryavcev.
\newblock Applying map-reduce paradigm for parallel closed cube computation.
\newblock In {\em 1st Int. Conf. on Advances in Databases, Knowledge, and Data
  Applications, {DBKDS} 2009}, pages 62--67, 2009.

\bibitem{Li:2009}
Ao~Li and David Tuck.
\newblock An effective tri-clustering algorithm combining expression data with
  gene regulation information.
\newblock {\em Gene regul. and syst. biol.}, 3:49--64, 2009.

\bibitem{Madeira:2004}
Sara~C. Madeira and Arlindo~L. Oliveira.
\newblock Biclustering algorithms for biological data analysis: A survey.
\newblock {\em IEEE/ACM Trans. Comput. Biology Bioinform.}, 1(1):24--45, 2004.

\bibitem{Metzler:2015}
Saskia Metzler and Pauli Miettinen.
\newblock Clustering boolean tensors.
\newblock {\em Data Min. Knowl. Discov.}, 29(5):1343--1373, 2015.

\bibitem{Mirkin:1996}
Boris Mirkin.
\newblock {\em Mathematical Classification and Clustering}.
\newblock Kluwer, Dordrecht, 1996.

\bibitem{Mirkin:2011}
Boris~G. Mirkin and Andrey~V. Kramarenko.
\newblock Approximate bicluster and tricluster boxes in the analysis of binary
  data.
\newblock In Sergei~O. Kuznetsov and et~al., editors, {\em RSFDGrC 2011},
  volume 6743 of {\em Lecture Notes in Computer Science}, pages 248--256.
  Springer, 2011.

\bibitem{Missaoui:2011}
Rokia Missaoui and L{\'{e}}onard Kwuida.
\newblock Mining triadic association rules from ternary relations.
\newblock In {\em Formal Concept Analysis - 9th International Conference,
  {ICFCA} 2011, Nicosia, Cyprus, May 2-6, 2011. Proceedings}, pages 204--218,
  2011.

\bibitem{Nanopoulos:2010}
Alexandros Nanopoulos, Dimitrios Rafailidis, Panagiotis Symeonidis, and Yannis
  Manolopoulos.
\newblock Musicbox: Personalized music recommendation based on cubic analysis
  of social tags.
\newblock {\em IEEE Transactions on Audio, Speech {\&} Language Processing},
  18(2):407--412, 2010.

\bibitem{PANCHENKO18}
Alexander Panchenko, Eugen Ruppert, Stefano Faralli, Simone~Paolo Ponzetto, and
  Chris Biemann.
\newblock {Building a Web-Scale Dependency-Parsed Corpus from CommonCrawl}.
\newblock In Nicoletta Calzolari~(Conference chair), Khalid Choukri,
  Christopher Cieri, Thierry Declerck, Sara Goggi, Koiti Hasida, Hitoshi
  Isahara, Bente Maegaard, Joseph Mariani, Hélène Mazo, Asuncion Moreno, Jan
  Odijk, Stelios Piperidis, and Takenobu Tokunaga, editors, {\em Proceedings of
  the Eleventh International Conference on Language Resources and Evaluation
  (LREC 2018)}, Miyazaki, Japan, May 7-12, 2018 2018. European Language
  Resources Association (ELRA).

\bibitem{Pawlak:1981}
Z.~Pawlak.
\newblock Information systems -- theoretical foundations.
\newblock {\em Information Systems}, 6(3):205 -- 218, 1981.

\bibitem{Rajaraman:2013}
Anand Rajaraman, Jure Leskovec, and Jeffrey~D. Ullman.
\newblock {\em Mining of Massive Datasets}, chapter MapReduce and the New
  Software Stack, pages 19--70.
\newblock Cambridge University Press, England, Cambridge, 2013.

\bibitem{Schweikardt18a}
Nicole Schweikardt.
\newblock One-pass algorithm.
\newblock In {\em Encyclopedia of Database Systems, Second Edition}. 2018.

\bibitem{Shin:2018}
Kijung Shin, Bryan Hooi, and Christos Faloutsos.
\newblock Fast, accurate, and flexible algorithms for dense subtensor mining.
\newblock {\em {TKDD}}, 12(3):28:1--28:30, 2018.

\bibitem{Spyropoulou:2014}
Eirini Spyropoulou, Tijl De~Bie, and Mario Boley.
\newblock Interesting pattern mining in multi-relational data.
\newblock {\em Data Mining and Knowledge Discovery}, 28(3):808--849, 2014.

\bibitem{Ustalov:2018}
Dmitry Ustalov, Alexander Panchenko, Andrey Kutuzov, Chris Biemann, and
  Simone~Paolo Ponzetto.
\newblock Unsupervised semantic frame induction using triclustering.
\newblock In Iryna Gurevych and Yusuke Miyao, editors, {\em Proceedings of the
  56th Annual Meeting of the Association for Computational Linguistics, {ACL}
  2018, Melbourne, Australia, July 15-20, 2018, Volume 2: Short Papers}, pages
  55--62. Association for Computational Linguistics, 2018.

\bibitem{Voutsadakis:02}
George Voutsadakis.
\newblock Polyadic concept analysis.
\newblock {\em Order}, 19(3):295--304, 2002.

\bibitem{Wille:1982}
Rudolf Wille.
\newblock Restructuring lattice theory: An approach based on hierarchies of
  concepts.
\newblock In Ivan Rival, editor, {\em Ordered Sets}, volume~83 of {\em NATO
  Advanced Study Institutes Series}, pages 445--470. Springer Netherlands,
  1982.

\bibitem{Xu:2012}
Biao Xu, Ruairı de~Frein, Eric Robson, and Mıcheal~O Foghlu.
\newblock Distributed formal concept analysis algorithms based on an iterative
  mapreduce framework.
\newblock In F.~Domenach, D.I. Ignatov, and J.~Poelmans, editors, {\em ICFCA
  2012}, volume LNAI 7278, pages 292--308, 2012.

\bibitem{Zaharia:2016}
Matei Zaharia, Reynold~S. Xin, Patrick Wendell, Tathagata Das, Michael
  Armbrust, Ankur Dave, Xiangrui Meng, Josh Rosen, Shivaram Venkataraman,
  Michael~J. Franklin, Ali Ghodsi, Joseph Gonzalez, Scott Shenker, and Ion
  Stoica.
\newblock Apache spark: A unified engine for big data processing.
\newblock {\em Commun. ACM}, 59(11):56–65, October 2016.

\bibitem{Zaki:2005}
Lizhuang Zhao and Mohammed~Javeed Zaki.
\newblock Tricluster: An effective algorithm for mining coherent clusters in 3d
  microarray data.
\newblock In {\em SIGMOD 2005 Conference}, pages 694--705, 2005.

\bibitem{Zudin:2015}
Sergey Zudin, Dmitry~V. Gnatyshak, and Dmitry~I. Ignatov.
\newblock Putting oac-triclustering on mapreduce.
\newblock In Sadok~Ben Yahia and Jan Konecny, editors, {\em Proceedings of the
  Twelfth International Conference on Concept Lattices and Their Applications,
  Clermont-Ferrand, France, October 13-16, 2015.}, volume 1466 of {\em {CEUR}
  Workshop Proceedings}, pages 47--58. CEUR-WS.org, 2015.

\end{thebibliography}

\end{document}